\documentclass[sigconf]{acmart}

\usepackage{balance}

\usepackage{stfloats} 
\usepackage{afterpage}
\usepackage{placeins}  

\usepackage{booktabs}
\usepackage{multirow}
\usepackage{arydshln}
\usepackage{xcolor}
\usepackage{float}
\usepackage{subfigure}
\usepackage{hyperref}

\AtBeginDocument{%
  }

\setcopyright{acmlicensed}
\copyrightyear{2018}
\acmYear{2018}
\acmDOI{XXXXXXX.XXXXXXX}

\acmConference[WWW '2025]{International World Wide Web Conference}{28 April-2 May 2025}{Sydney, Australia}
\acmISBN{XXXXXXX.XXXXXXX}

\begin{document}

\title{Scalable and Effective Negative Sample Generation for Hyperedge Prediction }



\author{Shilin Qu}
\email{shilin.qu@monash.edu}
\affiliation{%
  \institution{Monash University}
  \city{Melbourne}
  \country{Australia}
}

\author{Weiqing Wang}
\authornotemark[1]
\email{teresa.wang@monash.edu}
\affiliation{%
  \institution{Monash University}
  \city{Melbourne}
  \country{Australia}
}

\author{Yuan-Fang Li}
\authornotemark[1]
\email{yuanfang.li@monash.edu}
\affiliation{%
  \institution{Monash University}
  \city{Melbourne}
  \country{Australia}
}

\author{Quoc Viet Hung Nguyen}
\email{quocviethung.nguyen@griffith.edu.au}
\affiliation{%
  \institution{Griffith University}
  \city{Brisbane}
  \country{Australia}
}

\author{Hongzhi Yin}
\email{h.yin1@uq.edu.au}
\affiliation{%
  \institution{The University of Queensland}
  \city{Brisbane}
  \country{Australia}
}

\renewcommand{\shortauthors}{Shilin et al.}


\begin{abstract}
\label{sec:abstract}

Hyperedge prediction is crucial in hypergraph analysis for understanding complex multi-entity interactions in various web-based applications, including social networks and e-commerce systems. Traditional methods often face difficulties in generating high-quality negative samples due to the imbalance between positive and negative instances. To address this, we present the Scalable and Effective Negative Sample Generation for Hyperedge Prediction (SEHP) framework, which utilizes diffusion models to tackle these challenges. SEHP employs a boundary-aware loss function that iteratively refines negative samples, moving them closer to decision boundaries to improve classification performance. SEHP samples positive instances to form sub-hypergraphs for scalable batch processing. By using structural information from sub-hypergraphs as conditions within the diffusion process, SEHP effectively captures global patterns. To enhance efficiency, our approach operates directly in latent space, avoiding the need for discrete ID generation and resulting in significant speed improvements while preserving accuracy. Extensive experiments show that SEHP outperforms existing methods in accuracy, efficiency, and scalability, representing a substantial advancement in hyperedge prediction techniques. Our code is available here~\footnote{https://anonymous.4open.science/r/SEHP-09CD}.

\end{abstract}

\ccsdesc[300]{Information systems~Web log analysis}
\ccsdesc[500]{Information systems~Social networks}

\keywords{Hypergraph representation, Hyperedge Prediction, Negative Sample Generation, Conditional Diffusion}


\maketitle

\section{Introduction}
\label{sec:introduction}
 Graph models typically represent binary relationships, limiting their ability to capture complex multi-entity interactions~\cite{liu2020web,lin2021graph}. 
In contrast, hypergraphs connect multiple nodes through hyperedges, providing a flexible framework for modeling and integrating such interactions~\cite{unignn,hds,hypergcn,hgnn}. This makes hypergraphs particularly well-suited for web-based applications, including social networks~\cite{yu2021self,sun2021multi,liu2023disentangled}, collaborative networks~\cite{sun2023self}, e-commerce systems~\cite{han2023search,cheng2022ihgnn}, knowledge graph~\cite{wu2018approach,han2020open}, and web services~\cite{yahyaoui2021measuring,jia2023spatial},  where multi-dimensional relationships are prevalent. 

Hyperedge prediction, a fundamental task in hypergraph analysis, estimates the likelihood of potential connections and plays a critical role in these web-based applications. It supports tasks such as predicting community formation in social networks~\cite{sun2021multi,yu2021self}, identifying co-authorship in research networks~\cite{yoon2020much,wu2022hypergraph}, and detecting frequently co-purchased items in e-commerce~\cite{yang2022multi,zhang2021double}. 
A key aspect of hyperedge prediction is the vast number of potential negative examples, leading to a significant imbalance between positive and negative instances. For example, in the co-author network dataset Cora~\cite{hypergcn}, there are $2388$ nodes and $970$ observed positive hyperedges, resulting in approximately $(2^{2388} - 970)$ possible negative examples. This imbalance has motivated research efforts to sample high-quality negative examples for effective model training.

Most existing studies use rule-based negative sampling methods. For instance, HyperSAGNN~\cite{hypersagnn} samples negative examples as five times the positive samples based on a predefined noisy distribution~\cite{mikolov2013distributed,tu2018structural} to provide a robust learning signal. NHP~\cite{nhp} generates negative hyperedges by keeping half of the nodes from a positive hyperedge and randomly selecting the rest. However, these methods struggle with generalization due to their dependence on fixed sampling schemes. To overcome this, AHP~\cite{ahp} introduced a generative adversarial framework for negative hyperedge generation, setting a new benchmark in the field. Our work builds on this direction, further enhancing effectiveness and scalability.

Diffusion models~\cite{saharia2022photorealistic,austin2021structured,niu2020permutation}, known for their ability to enhance sample quality through iterative denoising, offer a promising solution for generating high-quality negative samples in hyperedge prediction. However, adapting these models to this task poses two key difficulties: (1) defining effective negative samples, as diffusion models are typically designed to generate positive samples which are usually clearly defined; and (2) adapting from continuous space where typical diffusion model operates in to discrete space where typical hyperedge prediction requires sampling in. Additionally, the poor scalability of current hyperedge prediction methods, such as NHP~\cite{nhp} and AHP~\cite{ahp}, remains a major challenge. These methods often require processing the entire hypergraph, resulting in high computational overhead. Such limitations are particularly critical for large-scale web-based systems, where both scalability and precise semantic processing are essential.

To address the challenges, we propose the \textbf{S}calable and \textbf{E}ffective Negative Sample Generation for \textbf{H}yperedge \textbf{P}rediction (\textbf{SEHP}) framework. 
In SEHP, we design a boundary-aware loss function that guides the diffusion model in generating negative hyperedges iteratively. This process ensures that each sample moves closer to the decision boundary, refining the quality of negative samples—essential for improving training outcomes in binary classification tasks like hyperedge prediction~\cite{dixit2023sampling,mackay1992information}. Different from existing hyperedge prediction methods, SEHP samples positive instances to form sub-hypergraphs for scalable batch processing. However, how to generate effective negative samples for the sampled sub-hypergraphs remains a challenge that has not been investigated yet. SEHP incorporates a conditional module that integrates structural information of the sub-hypergraph, ensuring the negative sample generating is also informed by broader global patterns. For adapting continuous space to discrete space, we attempted to map the continuous space to discrete space using a node ID extraction component. However, this approach was inefficient on GPUs due to varying node counts. To overcome this bottleneck, we developed an architecture that directly uses latent space representations, bypassing the need for discrete ID generation. This modification significantly accelerates the negative sample generation process, achieving a 20 to 71-fold speedup across six datasets while maintaining comparable performance with only a minor decrease (0.59\% in AUROC and 0.46\% in Precision). 

To summarize, our contributions are as follows:
\begin{itemize}
    \item We introduce \textbf{SEHP}, a novel framework for hyperedge prediction that leverages scalable and effective negative sample generation using conditional diffusion models, marking the first use of diffusion models in the field of hyperedge prediction.
    \item We design a boundary-aware loss function which defines effective negative samples to iteratively refine the negative sample generation process. 
    SEHP is the first batch-based hyperedge prediction method to effectively address the challenge of negative sample generation for sub-hypergraphs, using a conditional diffusion module that integrates global patterns. 
    Additionally, we investigate and demonstrate that generating negative samples directly from latent space is far more efficient than from discrete ID space, with only minimal performance loss.
    
    \item We perform extensive experiments, showing that \textbf{SEHP} outperforms existing state-of-the-art methods in terms of accuracy, efficiency, and scalability across multiple datasets.
\end{itemize}


\section{Related Work}
\label{sec:related_work}

Hypergraph neural networks (HGNNs)~\cite{hds,unignn,hgnn,yoon2020much} model complex multi-entity relationships. Unlike traditional GNNs~\cite{graphsage} focusing on pairwise links, HGNNs employ hyperedges for higher-order interactions. Methods like HyperGCN~\cite{hypergcn} (spectral approach) and UniGCNII~\cite{unignn} (unified message passing) enhance HGNN learning, while EDGNN~\cite{edgnn} uses equivariant diffusion to maintain node symmetry. However, these approaches mainly address node representation, neglecting hyperedge prediction challenges.

Hyperedge prediction is vital in areas like social networks, scientific collaboration, and pharmaceutical safety, focusing on identifying potential hyperedges. Techniques include link proximity ranking~\cite{10.1145/2487788.2487802}, matrix completion on incidence matrices~\cite{10.1145/3206025.3206062,10.5555/3294996.3295127}, hyperpath methods~\cite{10.1145/3357384.3357871}, and Laplacian tensor approaches for context-aware recommendations~\cite{10.1145/3269206.3269274}. These are mostly restricted to uniform hypergraphs. For non-uniform cases, Coordinated Matrix Minimization (CMM)~\cite{10.5555/3504035.3504578} uses non-negative matrix factorization and least squares in vertex adjacency space. Other studies examine small hyperlink evolution~\cite{Benson-2018-simplicial}. Recent machine learning advancements, such as HyperSAGNN~\cite{hypersagnn}, incorporates self-attention and higher negative sampling rates, and NHP~\cite{nhp}, a GCN-based model ranking hyperedges for better accuracy. AHP~\cite{ahp} further refines this with a generative adversarial framework for quality negative samples.


Negative sampling is critical in hyperedge prediction~\cite{nshpn2020, ahp, nhp}, affecting the model's ability to distinguish between positive and negative hyperedges. Traditional methods use heuristic techniques~\cite{nshpn2020, nhp, hypersagnn, 10.1145/1835804.1835837}, like randomly selecting \(k\) nodes or replacing half the nodes in a positive hyperedge, but they often lack generalization. An alternative is to learn negative sample generation. AHP~\cite{ahp}, for instance, uses a 3-layer MLP with LeakyReLU activations. However, AHP's approach is limited as it considers all nodes as candidates, which is impractical for large hypergraphs, and its generator uses only random noise, ignoring hypergraph structure, thus limiting performance.

Diffusion models are effective in generative tasks, perturbing data with Gaussian noise and training a reverse process to restore it. In images, Palette~\cite{saharia2022palette} uses conditional diffusion for colorization and inpainting, while Imagen~\cite{saharia2022photorealistic} extends this to text-to-image generation. 
EDP-GNN~\cite{niu2020permutation} models undirected graphs via score-matching diffusion on adjacency matrices. Graph Diffusion Convolution~\cite{gasteiger2019diffusion} uses localized convolution, leveraging heat kernel and personalized PageRank to manage noisy edges. 
While diffusion models focus on generating positive samples, none have explored their application in hyperedge prediction, especially for negative sample generation. Our work addresses this by applying diffusion models to generate negative samples for hyperedge prediction.


\section{PRELIMINARIES}
This section introduces the fundamental concepts and notations used in this paper. We define hypergraph elements, provide an overview of hyperedge prediction, and describe classical methods for negative hyperedge sampling.



\textit{Definition 1. (Hypergraph)}
A hypergraph \(\mathcal{G} = (\mathcal{V}, \mathcal{E})\) includes a set of nodes \(\mathcal{V}\) and hyperedges \(\mathcal{E}\), where each hyperedge \(e \in \mathcal{E}\) is a subset of \(\mathcal{V}\). It can be represented by an incidence matrix \(\mathbf{A} \in \{0, 1\}^{n \times m}\), where \(n = |\mathcal{V}|\) and \(m = |\mathcal{E}|\). \(\mathbf{A}_{i,j}\) is 1 if node \(i\) is in hyperedge \(j\), otherwise 0.

\textit{Definition 2. (Hyperedge Prediction)} 
Given a hypergraph \(\mathcal{G} = (\mathcal{V}, \mathcal{E})\) and node features \(\mathbf{X} \in \mathbb{R}^{n \times d}\), hyperedge prediction aims to identify a target set \(\mathcal{E}'\) where \(\mathcal{E}' \cap \mathcal{E} = \emptyset\). Each hyperedge \(e \in \mathcal{E}'\) is currently unobserved but expected to form in the future. The task can be formalized as a binary classification problem: for a candidate hyperedge \(e\), the goal is to predict whether \(e\) belongs to \(\mathcal{E}'\).

\subsection{Pipeline of Hyperedge Prediction}
\label{sec:Pipeline of Hyperedge Prediction}
The hyperedge prediction process consists of three main components~\cite{ahp,nhp,hypersagnn,nshpn2020}: \textit{Encoder}, \textit{Aggregator}, and \textit{Classifier}. First, the \textit{Encoder} transforms hypergraph data into node representations. Next, the \textit{Aggregator} learns candidate hyperedge representations from these node representations. Finally, the \textit{Classifier} predicts the existence of the candidate hyperedges based on the learned hyperedge representations.

\textbf{Encoder:} 
The hypergraph \textit{Encoder} focuses on learning node representations. The input consists of the hypergraph structure information $\mathbf{A}$ and node features $\mathbf{X}$. The output of the \textit{Encoder} is node embeddings. The formal definition of the \textit{Encoder} is as follows:
\begin{equation}
    \mathbf{V} = \text{HGNN}(\mathbf{A}, \mathbf{X})
    \label{eq:encoder}
\end{equation}
where \(\mathbf{V} \in \mathbb{R}^{n \times h}\) denotes the node embeddings, and $h$ is the embedding dimension. The \textit{Encoder} (\(\text{HGNN}\)) can be any hypergraph neural network, such as HyperGCN~\cite{hypergcn}, HyperSAGE~\cite{hypersage}, UniGNN~\cite{unignn}, EDGNN~\cite{edgnn}, or HDS~\cite{hds}.

\begin{figure*}[htbp]
    \centerline{\includegraphics[width=0.75\textwidth]{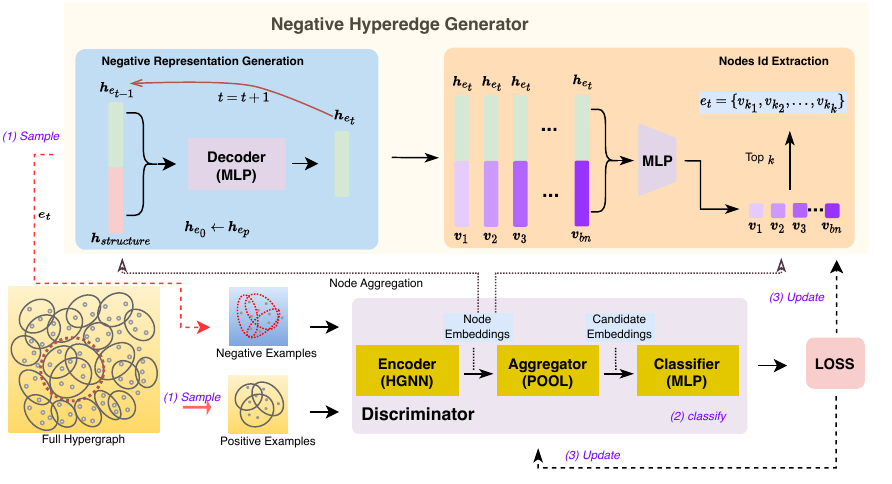}}
    \caption{Overview of the Conditional Diffusion-Based Framework for Negative Hyperedge Generation in Hyperedge Prediction. The framework consists of three main components: (1) sampling from the full hypergraph to generate positive samples for batch processing; (2) a discriminator module with an encoder, aggregator, and classifier that evaluates candidate hyperedges; (3) a negative hyperedge generator utilizing conditional diffusion and node ID extraction to create challenging negative samples for training. The iterative process updates the model based on the loss function to refine prediction accuracy.}
    \Description{Overview of the Conditional Diffusion-Based Framework for Negative Hyperedge Generation in Hyperedge Prediction. The framework consists of three main components: (1) sampling from the full hypergraph to generate positive samples; (2) a discriminator module with an encoder, aggregator, and classifier that evaluates candidate hyperedges; (3) a negative hyperedge generator utilizing conditional diffusion and node ID extraction to create challenging negative samples for training. The iterative process updates the model based on the loss function to refine prediction accuracy.}
    \label{fig:framework}
\end{figure*}

\textbf{Aggregator:}
The \textit{Aggregator} focuses on obtaining the representation of a hyperedge based on the learned node representation from the \textit{Encoder}~\cite{Benson-2018-simplicial,ahp}. The input of the \textit{Aggregator} is a set of node embeddings learned by the \textit{Encoder}, and the output is a candidate hyperedge embedding. The formal definition of the \textit{Aggregator} is as follows:
\begin{equation}
    \mathbf{e}_c = \text{Aggregator}(\mathbf{V}_{c})
    \label{eq:aggregator}
\end{equation}
where \(\mathbf{V}_{c} \in \mathbb{R}^{|e_c| \times h}\) represents the embeddings of nodes in the candidate hyperedge \(e_c\), and \(\mathbf{e}_c \in \mathbb{R}^{h}\) is the candidate hyperedge embedding. The \textit{Aggregator} can be any hypergraph aggregation method, such as maxmin, sum, mean, max, or attention-based aggregation~\cite{ahp,nhp,hypersagnn}.

\textbf{Classifier:} 
The \textit{Classifier} predicts the existence of candidate hyperedges~\cite{Benson-2018-simplicial,ahp,nhp}. It takes the candidate hyperedge embedding learned by the \textit{Aggregator} as input and outputs a probability score \(s_{c} \in [0, 1]\). The formal definition of the \textit{Classifier} is as follows:
\begin{equation}
    s_{c} = \text{Classifier}(\mathbf{e}_c)
    \label{eq:classifier}
\end{equation}
The \textit{Classifier} can be any neural network, such as a feedforward neural network, as long as it can output a probability score.

\subsection{Negative Hyperedge Sampling}
\label{sec:Negative Hyperedge Sampling preliminaries}
Negative hyperedge sampling is essential for training effective hyperedge prediction models, as it enhances performance and generalization~\cite{nshpn2020,ahp,nhp}. The process selects negative hyperedges not in the observed set \(\mathcal{E}\), aiming to sample true negatives that are not part of the future set \(\mathcal{E}'\).

Negative hyperedges are sampled from the unobserved set \(2^{|\mathcal{V}|} \setminus \mathcal{E}\) using various heuristic approaches such as sized negative sampling (SNS)~\cite{hypersagnn,ahp}, motif negative sampling (MNS)~\cite{10.1109/ICDM.2012.87}, and clique negative sampling (CNS)~\cite{10.1145/1835804.1835837, nshpn2020}. In SNS, \(k\) nodes are randomly selected from the hypergraph without considering its structure. MNS begins with an observed hyperedge and expands by merging random incident edges until it reaches size \(k\), taking into account local connectivity patterns. CNS selects an existing hyperedge and replaces one node with an adjacent node, preserving the original structure while introducing slight variations.

Generative sampling methods, such as AHP~\cite{ahp}, employ an adversarial framework to generate negative hyperedges. AHP, for example, uses an MLP with LeakyReLU~\cite{maas2013rectifier} activations but relies solely on random noise as input, disconnected from the hypergraph structure, and the output matches the total number of nodes. Consequently, AHP struggles with generating high-quality samples and scaling to larger hypergraphs. To address these limitations, we propose a conditional diffusion model that selects candidate nodes from the current batch sub-hypergraph, improving both scalability and sample quality.

\section{Proposed Method}
In this section, we present our method for scalable and effective hyperedge prediction. We first build a hyperedge discriminator. Next, we develop a conditional diffusion-based negative hyperedge generator with batch processing. Finally, we outline the loss function and acceleration techniques used to optimize the framework.

\subsection{Discriminator}
Given a subset of nodes, the discriminator determines whether they can form hyperedges. It comprises three components: an encoder, an aggregator, and a classifier, as outlined in Section~\ref{sec:Pipeline of Hyperedge Prediction}. The encoder converts hypergraph data into node embeddings, the aggregator combines these embeddings to form candidate hyperedge representations, and the classifier predicts the likelihood of the candidate hyperedges existing. Our implementation is based on PyTorch-Geometric\footnote{\url{https://pytorch-geometric.readthedocs.io}}. For the encoder, we represent the hypergraph as a bipartite graph, a common technique in many hypergraph methods~\cite{wan2021principled,10.1145/3457949}. In this bipartite representation, the original nodes of the hypergraph form one set, while the hyperedges form the other. We use GraphSAGE~\cite{graphsage} to learn node embeddings for this bipartite graph structure. The aggregator component employs a max-min pooling, shown to be effective in prior work~\cite{nhp}, to generate candidate hyperedge embeddings. Finally, the classifier is implemented as a multi-layer perceptron (MLP) with a sigmoid activation function, following previous studies~\cite{ahp,nhp}, to output the probability of the candidate hyperedge's existence.

\subsection{Negative Hyperedge Generator with Batch Processing}
\label{sec:negativeSampling}
The primary goal of the generator in hyperedge prediction is to produce high-quality negative samples, which differs significantly from conventional generators used in fields like computer vision~\cite{10447191,Chen_2023_ICCV,pmlr-v202-bao23a} and NLP~\cite{NEURIPS2022_1be5bc25,he-etal-2023-diffusionbert,NEURIPS2023_35b5c175}, where the focus is on generating high-quality positive samples. If the generator is too weak, it produces only trivial negative samples, which are ineffective for training the discriminator. On the other hand, if the generator is too strong and starts producing positive hyperedge samples similar to those in the target set, it fails to provide true negative samples.

Diffusion models have showcased substantial capabilities in sample generation. Nonetheless, adapting these models for negative hyperedge generation presents distinct challenges: (1) \textbf{Defining an effective ``good'' negative sample is problematic as diffusion models are fundamentally designed to produce well-defined, high-quality positive samples, rather than negative ones}; (2) \textbf{Diffusion models operate in a continuous space, while negative sampling for this task necessitates operations in a discrete space}. For scalability, the diffusion process should be constrained to the batched sub-hypergraph space, sampled as described in Section~\ref{Positive Hyperedges Sampling for Batch Processing}. However, as training progresses, the composition of the sub-hypergraph varies with each batch change. (3) \textbf{This variability in sub-hypergraph composition is not inherently recognized by diffusion models, potentially impacting the quality of generated negative samples}.

In the following subsections, we present strategies to integrate diffusion models effectively into the negative hyperedge generation process, addressing these challenges to enhance model performance and scalability.

\subsubsection{Positive Hyperedges Sampling for Batch Processing}
\label{Positive Hyperedges Sampling for Batch Processing}
To enable the scalability of the hyperedge prediction model, we sample sub-hypergraphs for batch training. While various sampling methods exist, such as random sampling, node-based sampling, and hyperedge-based sampling~\cite{ yoon2020much}, we adopt a neighbor-based approach~\cite{nshpn2020} as it produces denser sub-hypergraphs. Starting with a randomly selected hyperedge, we iteratively expand to neighboring hyperedges—those sharing at least one node—until the sub-hypergraph reaches the desired size for batch processing.

\subsubsection{Diffusion-Based Negative Hyperedge Representation Generation}
Conventionally, diffusion methods~\cite{NEURIPS2020_4c5bcfec,NEURIPS2021_49ad23d1} consist of two phases: adding noise and denoising. First, noise is iteratively added to the original sample until it becomes random noise, thus creating labeled data for training. Then, the denoising process aims to iteratively recover the noised data to the original sample. However, in hyperedge prediction, there is no clear notion of what constitutes a good negative sample, so there is no original sample to start with. Instead, the score predicted by the classifier can be used as a signal to supervise the denoising process. Given a negative hyperedge representation at time $t$, a new negative hyperedge representation at time $t+1$ can be generated by denoising the current negative hyperedge representation. The signal for denoising is ${\text{score}_{\text{hyperedge}}^{t} < \text{score}_{\text{hyperedge}}^{t+1}}$, and the process of denoising is defined as follows:
\begin{equation}
    \mathbf{h}_{e}^{t+1} = \text{Denoise}(\mathbf{h}_{e}^{t})
    \label{eq:denoise}
\end{equation}
where $\mathbf{h}_{e}^{t}$ is the negative hyperedge representation at time $t$, and $\mathbf{h}_{e}^{t+1}$ is the negative hyperedge representation at time $t+1$. Here, $\mathbf{h}_{e}^{0} = \mathbf{h}_{\hat{e}}^{p}$, where $\hat{e}^{p}$ is a representation of a positive hyperedge $\hat{e}$. The denoising process can be any neural network, such as an MLP, recurrent neural network~\cite{SIP-2022-0023}, or graph neural network~\cite{graphsage}. In this work, an MLP is used for the denoising process.

\subsubsection{Conditional Denoising for Controlled Diffusion}
As training progresses, the composition of the sub-hypergraph changes with each batch, a variation not inherently accounted for by traditional diffusion models. To better control the diffusion process and adapt to these changes, we introduce a conditional mechanism during the denoising phase. This mechanism addresses the limitation that information from a single positive hyperedge is often insufficient for generating high-quality negative samples. Instead, we leverage the structural information of the sub-hypergraph, which encapsulates a more comprehensive view of the global data. By incorporating this structural information as an additional input in the denoising process, the conditional mechanism enhances the generation of negative hyperedge representations. Specifically, when combined with the candidate hyperedge representation, this integration facilitates the generation of improved negative samples that are more aligned with the global structure of the sub-hypergraph. The conditional denoising process is formally defined as follows:
\begin{equation}
    \mathbf{h}_{e}^{t} = \text{Denoise}(\mathbf{h}_{e}^{t-1}, \mathbf{h}_{structure}^{sub})
    \label{eq:conditional_denoise}
\end{equation}
where $\mathbf{h}_{structure}^{sub}$ represents the structural information of the sub-hypergraph, and $\mathbf{h}_{e}^{t}$ denotes the negative hyperedge representation at time $t$. The structural information, $\mathbf{h}_{structure}^{sub}$, is obtained by applying a readout function over the node embeddings of the sub-hypergraph. This readout can be performed using various methods, such as sum, mean, max, or attention-based techniques. In our approach, we employ an average readout process to capture the sub-hypergraph's structure. The entire process is illustrated in the component "Negative Representation Generation" of Figure~\ref{fig:framework}.


\subsubsection{Node ID Extraction}
In hyperedge prediction using diffusion models, converting a continuous representation into discrete node IDs is a challenge. While diffusion processes are effective in continuous spaces, hyperedge prediction requires the identification of specific nodes that form a negative hyperedge, necessitating discrete outputs. Therefore, mapping the continuous data generated by the diffusion model back into discrete node IDs is essential for constructing valid negative hyperedges.

We achieve this by combining negative hyperedge representations with candidate node embeddings from the sub-hypergraph.
Using a machine learning model, typically a multi-layer perceptron (MLP), we estimate the probability of each node being included in the negative hyperedge. The formal definition of the node ID extraction process is:
\begin{equation}
    e_t =\{\hat{v}_1^{sub}, \ldots, \hat{v}_k^{sub}\} = \text{Topk}(p_{v_1}, \ldots, p_{v_n})
    \label{eq:node_id_extraction}
\end{equation}
where $p_{v_i} = \text{MLP}(\mathbf{h}_{e}^t, \mathbf{v}_i)$, and $e_t$ denotes the negative hyperedge at time $t$. Here, $\hat{v}_i^{sub}$ represents the $i$-th node in the negative hyperedge, and $p_{v_i}$ is the probability of node $v_i$ being part of the negative hyperedge, as determined by the MLP output. The MLP is equipped with a sigmoid activation function to ensure the output probabilities are within the range (0, 1).

This process not only converts the continuous representations generated by the diffusion model into a set of discrete node IDs but also selects the most probable candidates to form the hyperedge based on their computed probabilities. By doing so, this method effectively bridges the gap between the generative capabilities of diffusion models and the discrete requirements of hyperedge prediction tasks, ultimately enhancing the system's overall performance and accuracy.

\subsection{Loss Function}
The SEHP loss function is designed to guide the training of both the discriminator and the negative hyperedge generator, ensuring that the generator produces challenging negative samples while the discriminator correctly distinguishes between positive and negative instances. The discriminator's loss is a binary cross-entropy loss:

\begin{equation}
    \mathcal{L}_{dis} = -\frac{1}{N} \sum_{i=1}^{N} \left[ y_i \log(s_i) + (1-y_i) \log(1-s_i) \right]
    \label{eq:loss_dis}
\end{equation}
where $N$ is the number of samples, $y_i \in \{0,1\}$ denotes the ground truth label, and $0 \leq s_i \leq 1$ is the predicted probability score, ensuring that the discriminator correctly classifies the samples.

For the negative hyperedge generator, the loss consists of two components. The first component, $\mathcal{L}_{neg}$, encourages the generator to create negative samples that are challenging for the discriminator:
\begin{equation}
    \mathcal{L}_{neg} = -\frac{1}{N} \sum_{i=1}^{N} s_i
    \label{eq:loss_neg}
\end{equation}
The second component, $\mathcal{L}_{diff}$, which is the boundary-aware loss function, ensures that the generated negative samples move closer to the decision boundary through the iterative denoising process:
\begin{equation}
    \mathcal{L}_{diff} = \log \left( \frac{s_i^{t-1}}{s_i^t} \right)
    \label{eq:loss_diff}
\end{equation}
The total loss for the generator combines these two components:
\begin{equation}
    \mathcal{L}_{gen} = \mathcal{L}_{neg} + \mathcal{L}_{diff}
    \label{eq:loss_gen_final}
\end{equation}

The losses for the discriminator and generator are optimized alternately during training, ensuring that both components improve simultaneously, resulting in a balanced and effective model for hyperedge prediction.

\subsection{Accelerating Negative Hyperedge Generation}
\label{sec:accelerating_training_process}
Although our framework is already more scalable than existing hyperedge prediction methods due to its use of sub-hypergraphs, the Node ID extraction process remains a bottleneck when handling large hypergraphs. This process, being discrete, must be executed separately for each negative hyperedge, leading to inefficiencies, especially when hyperedges have different node counts (e.g., $\hat{e}_1$ with $k_{e_1}$ nodes and $\hat{e}_2$ with $k_{e_2}$ nodes where $k_{e_1} \neq k_{e_2}$). Despite the uniformity in the shapes of their generated negative representations, the differences in node counts necessitate separate extraction operations, unlike other components, such as Negative Representation Generation and the discriminator, which can be efficiently parallelized on a GPU.

To address this inefficiency, we propose using the continuous negative hyperedge representation generated by the diffusion process directly, bypassing the need for discrete Node ID extraction~\cite{pmlr-v139-nichol21a,10.1145/3631116}. This approach, as illustrated in Figure~\ref{fig:framework_edge_representation}, allows the generated representation to serve directly as the negative hyperedge embedding. The classifier then utilizes this embedding to predict the probability of the candidate hyperedge, as indicated by the direction of the purple arrow in Figure~\ref{fig:framework_edge_representation} in Appendix. 

By eliminating the Node ID extraction step, we overcome the discrete computation challenge, significantly improving the efficiency of the training process. The loss function for the negative hyperedge generator remains unchanged, but the streamlined approach accelerates training. The experimental results (see Table~\ref{tab:hyperedge_prediction_ablation}) demonstrate the effectiveness of this method, showing the improvements in speed and performance when using the continuous representation directly.


\section{Experiment}

\begin{table*}[htbp]
    \caption{Results of hyperedge prediction on four small datasets (Cora, DBLP, Citeseer, and NDC\_class), showing AUROC and Precision metrics for various baselines and the proposed SEHP model across different test sets (SNS, MNS, CNS, and MIX).}

    \centering
    \resizebox{0.98 \textwidth}{!}{%
    \begin{tabular}{c|cccccccccc|cccccccccc}
        \hline
        Dataset  & \multicolumn{10}{c|}{Cora} & \multicolumn{10}{c}{DBLP} \\ \hline
        Matrices & \multicolumn{5}{c|}{AUROC} & \multicolumn{5}{c|}{Precision} & \multicolumn{5}{c|}{AUROC} & \multicolumn{5}{c}{Precision}                  \\ \hline
        Test Set & SNS & MNS & CNS & \multicolumn{1}{c|}{MIX} & \multicolumn{1}{c|}{AVE} & SNS & MNS & CNS & \multicolumn{1}{c|}{MIX} & AVE & SNS & MNS & CNS & \multicolumn{1}{c|}{MIX} & \multicolumn{1}{c|}{AVE} & SNS & MNS & CNS & \multicolumn{1}{c|}{MIX} & AVE \\ \hline
        HyperGCN & 0.645 & 0.571 & 0.285 & \multicolumn{1}{c|}{0.497}    & \multicolumn{1}{c|}{0.501}    & 0.583 & 0.511 & 0.365 & \multicolumn{1}{c|}{0.494} & \multicolumn{1}{c|}{0.488}  & 0.571 & 0.595 & 0.307 & \multicolumn{1}{c|}{0.466}    & \multicolumn{1}{c|}{0.485}    & 0.535 & 0.543 & 0.398 & \multicolumn{1}{c|}{0.493} & \multicolumn{1}{c}{0.492}  \\ \hdashline[0.5pt/2pt]
        UniGCNII & 0.588 & 0.429 & 0.415 & \multicolumn{1}{c|}{0.420}    & \multicolumn{1}{c|}{0.463}    & 0.565 & 0.418 & 0.425 & \multicolumn{1}{c|}{0.429} & \multicolumn{1}{c|}{0.459}  & 0.564 & 0.361 & 0.351 & \multicolumn{1}{c|}{0.429}    & \multicolumn{1}{c|}{0.426}    & 0.551 & 0.424 & 0.399 & \multicolumn{1}{c|}{0.455} & \multicolumn{1}{c}{0.457}  \\ \hdashline[0.5pt/2pt]
        EDGNN & 0.574 & 0.189 & 0.388 & \multicolumn{1}{c|}{0.370}    & \multicolumn{1}{c|}{0.380}    & 0.584 & 0.388 & 0.422 & \multicolumn{1}{c|}{0.442} & \multicolumn{1}{c|}{0.459}  & 0.809 & 0.367 & 0.391 & \multicolumn{1}{c|}{0.526}    & \multicolumn{1}{c|}{0.523}    & 0.501 & 0.492 & 0.496 & \multicolumn{1}{c|}{0.501} & \multicolumn{1}{c}{0.497}  \\ \hdashline[0.5pt/2pt]
        HDS & 0.716 & 0.617 & 0.329 & \multicolumn{1}{c|}{0.562}    & \multicolumn{1}{c|}{0.556}    & 0.666 & 0.608 & 0.370 & \multicolumn{1}{c|}{0.535} & \multicolumn{1}{c|}{0.545}  & 0.714 & 0.554 & 0.389 & \multicolumn{1}{c|}{0.558}    & \multicolumn{1}{c|}{0.554}    & 0.638 & 0.512 & 0.431 & \multicolumn{1}{c|}{0.526} & \multicolumn{1}{c}{0.527}  \\ \hline
        HyperSAGNN  & 0.511 & 0.458 & 0.467 & \multicolumn{1}{c|}{0.514}    & \multicolumn{1}{c|}{0.488}    & 0.502 & 0.517 & 0.501 & \multicolumn{1}{c|}{0.533} & \multicolumn{1}{c|}{0.513}  & 0.955 & 0.687 & 0.590 & \multicolumn{1}{c|}{0.747}    & \multicolumn{1}{c|}{0.745}    & 0.502 & 0.501 & 0.561 & \multicolumn{1}{c|}{0.502} & \multicolumn{1}{c}{0.515}  \\ \hdashline[0.5pt/2pt]
        NHP & 0.532 & 0.289 & 0.239 & \multicolumn{1}{c|}{0.415}    & \multicolumn{1}{c|}{0.369}    & 0.527 & 0.317 & 0.293 & \multicolumn{1}{c|}{0.415} & \multicolumn{1}{c|}{0.388}  & 0.837 & 0.645 & 0.584 & \multicolumn{1}{c|}{0.714}    & \multicolumn{1}{c|}{0.695}    & 0.664 & 0.530 & 0.531 & \multicolumn{1}{c|}{0.577} & \multicolumn{1}{c}{0.575}  \\ \hdashline[0.5pt/2pt]
        AHP & 0.808 & 0.802 & 0.734 & \multicolumn{1}{c|}{0.714}    & \multicolumn{1}{c|}{0.765}    & 0.661 & 0.434 & 0.322 & \multicolumn{1}{c|}{0.262} & \multicolumn{1}{c|}{0.420}  & 0.751 & 0.765 & 0.692 & \multicolumn{1}{c|}{0.692}    & \multicolumn{1}{c|}{0.725}    & 0.780 & 0.659 & 0.414 & \multicolumn{1}{c|}{0.343} & \multicolumn{1}{c}{0.549}  \\ \hline
        SEHP(Ours) &
        \textbf{0.974} & \textbf{0.951} & \textbf{0.850} & \multicolumn{1}{c|}{\textbf{0.856}}    & \multicolumn{1}{c|}{\textbf{0.908}}    & \textbf{0.746} & \textbf{0.774} & \textbf{0.638} & \multicolumn{1}{c|}{\textbf{0.654}} & \multicolumn{1}{c|}{\textbf{0.703}} 
        & \textbf{0.961} & \textbf{0.827} & \textbf{0.749} &  \multicolumn{1}{c|}{\textbf{0.847}} &  \multicolumn{1}{c|}{\textbf{0.846}} &     \textbf{0.855} & \textbf{0.690} & \textbf{0.620} & \multicolumn{1}{c|}{\textbf{0.682}} & \multicolumn{1}{c}{\textbf{0.712}}   
        \\ \hline
    \end{tabular} 
    }
    \resizebox{0.98 \textwidth}{!}{%
    \begin{tabular}{c|cccccccccc|cccccccccc}
        \hline \hline
        Dataset  & \multicolumn{10}{c|}{Citeseer}& \multicolumn{10}{c}{NDC\_class}\\ \hline
        Matrices & \multicolumn{5}{c|}{AUROC} & \multicolumn{5}{c|}{Precision}  & \multicolumn{5}{c|}{AUROC}  & \multicolumn{5}{c}{Precision}\\ \hline
        Test Set & SNS & MNS & CNS & \multicolumn{1}{c|}{MIX} & \multicolumn{1}{c|}{AVE} & SNS & MNS & CNS & \multicolumn{1}{c|}{MIX} & AVE & SNS & MNS & CNS & \multicolumn{1}{c|}{MIX} & \multicolumn{1}{c|}{AVE} & SNS & MNS & CNS & \multicolumn{1}{c|}{MIX} & AVE \\ \hline
        HyperGCN & 0.814 & 0.579 & 0.286 & \multicolumn{1}{c|}{0.535}  & \multicolumn{1}{c|}{0.553}    & 0.722 & 0.489 & 0.359 & \multicolumn{1}{c|}{0.509} & \multicolumn{1}{c|}{0.5201}  & 0.778 & 0.490 & 0.235 & \multicolumn{1}{c|}{0.509}    & \multicolumn{1}{c|}{0.503}    & 0.677 & 0.475 & 0.313 & \multicolumn{1}{c|}{0.515} & \multicolumn{1}{c}{0.495}  \\ \hdashline[0.5pt/2pt]
        UniGCNII & 0.503 & 0.326 & 0.279 & \multicolumn{1}{c|}{0.395}    & \multicolumn{1}{c|}{0.375}    & 0.515 & 0.389 & 0.355 & \multicolumn{1}{c|}{0.435} & \multicolumn{1}{c|}{0.423}  & 0.805 & 0.585 & 0.486 & \multicolumn{1}{c|}{0.660}    & \multicolumn{1}{c|}{0.634}    & 0.734 & 0.526 & 0.493 & \multicolumn{1}{c|}{0.578} & \multicolumn{1}{c}{0.583}  \\ \hdashline[0.5pt/2pt]
        EDGNN & 0.752 & 0.301 & 0.251 & \multicolumn{1}{c|}{0.463}    & \multicolumn{1}{c|}{0.442}    & 0.510 & 0.417 & 0.383 & \multicolumn{1}{c|}{0.502} & \multicolumn{1}{c|}{0.451}  & 0.872 & 0.575 & 0.380 & \multicolumn{1}{c|}{0.569}    & \multicolumn{1}{c|}{0.599}    & 0.502 & 0.464 & 0.450 & \multicolumn{1}{c|}{0.545} & \multicolumn{1}{c}{0.490}  \\ \hdashline[0.5pt/2pt]
        HDS & 0.765 & 0.441 & 0.421 & \multicolumn{1}{c|}{0.521}    & \multicolumn{1}{c|}{0.537}    & \textbf{0.756} & 0.415 & 0.395 & \multicolumn{1}{c|}{0.468} & \multicolumn{1}{c|}{0.509}  & 0.829 & 0.567 & 0.325 & \multicolumn{1}{c|}{0.556}    & \multicolumn{1}{c|}{0.569}    & 0.501 & 0.419 & 0.401 & \multicolumn{1}{c|}{0.504} & \multicolumn{1}{c}{0.456}  \\ \hline
        HyperSAGNN  & 0.572 & 0.529 & 0.495 & \multicolumn{1}{c|}{0.517}    & \multicolumn{1}{c|}{0.528}    & 0.604 & 0.538 & 0.508 & \multicolumn{1}{c|}{0.516} & \multicolumn{1}{c|}{0.541}  & 0.701 & 0.572 & 0.601 & \multicolumn{1}{c|}{0.612}    & \multicolumn{1}{c|}{0.622}    & 0.829 & 0.669 & 0.640 & \multicolumn{1}{c|}{0.632} & \multicolumn{1}{c}{0.692}  \\ \hdashline[0.5pt/2pt]
        NHP & 0.734 & 0.437 & 0.344 & \multicolumn{1}{c|}{0.450}    & \multicolumn{1}{c|}{0.491}    & 0.742 & 0.462 & 0.369 & \multicolumn{1}{c|}{0.478} & \multicolumn{1}{c|}{0.513}   & 0.839 & 0.786 & 0.714  & \multicolumn{1}{c|}{0.721} & \multicolumn{1}{c|}{0.765} &  0.577 & 0.375 & 0.272  &  \multicolumn{1}{c|}{0.219} & \multicolumn{1}{c}{0.361}  \\ \hdashline[0.5pt/2pt]
        AHP & 0.628 & 0.662 & 0.631 & \multicolumn{1}{c|}{0.635}    & \multicolumn{1}{c|}{0.639}    & 0.661 & 0.519 & 0.358 & \multicolumn{1}{c|}{0.298} & \multicolumn{1}{c|}{0.459}  & 0.861 & 0.799 & 0.729 & \multicolumn{1}{c|}{0.725}    & \multicolumn{1}{c|}{0.779}    & 0.798 & 0.586 & 0.375 & \multicolumn{1}{c|}{0.304} & \multicolumn{1}{c}{0.516}  \\ \hline 
        SEHP (Ours) & \textbf{0.948} & \textbf{0.841} & \textbf{0.700} & \multicolumn{1}{c|}{\textbf{0.841}}    & \multicolumn{1}{c|}{\textbf{0.832}}    & 0.763 & \textbf{0.719} & \textbf{0.629} & \multicolumn{1}{c|}{\textbf{0.676}} & \multicolumn{1}{c|}{\textbf{0.686}}  & \textbf{0.982} & \textbf{0.861} & \textbf{0.788} & \multicolumn{1}{c|}{\textbf{0.873}}    & \multicolumn{1}{c|}{\textbf{0.876}}    & \textbf{0.850} & \textbf{0.771} & \textbf{0.665} & \multicolumn{1}{c|}{\textbf{0.721}} & \multicolumn{1}{c}{\textbf{0.752}}  \\ \hline
    \end{tabular}
    }
    \label{tab:hyperedge_prediction_baselines}
\end{table*}

We present experiments to evaluate the effectiveness of our model and its components in hyperedge prediction, addressing the following research questions (RQs).
\begin{itemize}
    \item \textbf{RQ1: Is SEHP more effective and scalable than existing models?} SEHP consistently outperformed state-of-the-art methods across four small datasets (Table~\ref{tab:hyperedge_prediction_baselines}). On two large datasets, existing models designed for hyperedge prediction failed to run due to scalability limitations, while SEHP demonstrated superior performance over all baselines (Table~\ref{tab:large_scale}).
    \item \textbf{RQ2: Does incorporating global structural information during the negative sample generation for sub-hypergraphs enhance hyperedge prediction, particularly for large hypergraphs?} An ablation study, conducted with and without structural information, confirmed the beneficial impact of incorporating structural information on performance (see Tables~\ref{tab:hyperedge_prediction_ablation} and \ref{tab:hyperedge_prediction_ablation_large}).
    \item \textbf{RQ3: Is the diffusion model effective in generating negative samples?} The effectiveness of the diffusion model was validated through an ablation study comparing the negative generator with and without diffusion. Furthermore, using negative representations directly in the classifier yielded competitive results (see Tables~\ref{tab:hyperedge_prediction_ablation} and \ref{tab:hyperedge_prediction_ablation_large}).
    \item \textbf{RQ4: Is the negative hyperedge generation acceleration scheme effective?} An ablation study compared  Node ID extraction method with our acceleration scheme, which generates samples directly from the latent space. The results indicate that our approach is 20 to 71 times faster across various datasets, with only a negligible performance trade-off (less than 1\% on average) (see Table~\ref{tab:training_time}).
\end{itemize}

\subsection{Datasets, Setting and Baselines}
\paragraph[short]{Datasets} We use six real-world datasets summarized in Table~\ref{tab:dataset_summary}: two authorship networks (Cora, DBLP), one co-citation network (Citeseer), one drug classification dataset (NDC\_class), and two recipe datasets (Recipe100k, Recipe200k). \textbf{Cora} and \textbf{DBLP}~\cite{hypergcn} represent papers and their co-authorships, with node features based on bag-of-words from abstracts. Cora contains 2,388 nodes and 970 hyperedges, while DBLP has 41,302 nodes and 20,865 hyperedges. \textbf{Citeseer}~\cite{hypergcn} is a co-citation network where nodes represent papers and hyperedges represent citation groups, with a total of 1,458 nodes and 1,004 hyperedges. \textbf{NDC\_class}~\cite{Benson-2018-simplicial} includes drugs as hyperedges and class labels as nodes, using one-hot encodings for node features, with 1,161 nodes and 1,047 hyperedges. \textbf{Recipe100k} and \textbf{Recipe200k}~\cite{li2022share} are recipe-ingredient networks that use bag-of-words features from recipe texts; Recipe100k contains 100,896 nodes and 11,822 hyperedges, while Recipe200k includes 240,094 nodes and 18,049 hyperedges.

\paragraph[short]{Setting} We report AUROC\footnote{\url{https://lightning.ai/docs/torchmetrics/stable/classification/auroc.html\#torchmetrics.classification.BinaryAUROC}} and Precision metrics, consistent with prior studies~\cite{ahp,nhp}. 
Each dataset is split into 60\% training, 20\% validation, and 20\% testing. Following previous work~\cite{ahp,nhp,hypersagnn}, we generate negative samples for the validation and test sets to calculate these metrics. For smaller datasets (Cora, DBLP, Citeseer, NDC\_class), we create four sets using SNS, MNS, CNS, and a mixed approach (MIX, as detailed in subsection~\ref{sec:Negative Hyperedge Sampling preliminaries}). For larger datasets (Recipe100k and Recipe200k), we use SNS to focus on scalability, as MNS and CNS are time-consuming. Test results are reported based on the highest AUROC and Precision values achieved on the validation sets.
\paragraph[short]{Baselines} We evaluate our approach against two groups of baselines for hyperedge prediction. The first group includes the state-of-the-art methods specifically designed for hyperedge prediction: \textbf{HyperSAGNN}~\cite{hypersagnn}, \textbf{NHP}~\cite{nhp}, and \textbf{AHP}~\cite{ahp}, as detailed in Section~\ref{sec:related_work}. The second group consists of hypergraph embedding methods adapted for hyperedge prediction, including \textbf{HyperGCN}~\cite{hypergcn}, \textbf{UniGCNII}~\cite{unignn}, \textbf{HDS}~\cite{hds}, and \textbf{EDGNN}~\cite{edgnn}. These methods are used as encoders, followed by an aggregator and a classifier, same to the structure employed in SEHP.



\subsection{Results}
Here, we first present the primary results for four small datasets and two large ones seperately. Following this, an ablation study will illustrate the efficacy of the essential elements in our approach. 
\subsubsection{Main Results}
\textbf{Compared with SOTA on Small Datasets.}
\label{sec:Compared with SOTA on Small Datasets}

As shown in Table~\ref{tab:hyperedge_prediction_baselines}, our model SEHP outperformed all state-of-the-art models on all 16 test sets for AUROC and on 13 for Precision. For instance, on the MIX test sets, SEHP consistently outperformed the second-best models across all datasets. On the Cora dataset, SEHP achieved an AUROC of 0.856, surpassing AHP's 0.714 by 19.89\%, and a Precision of 0.654, exceeding HDS's 0.535 by 22.24\%. Similarly, SEHP improved AUROC by 13.39\% and Precision by 18.19\% on DBLP, while showing improvements of 32.44\% (AUROC) and 31.01\% (Precision) on Citeseer, and 20.41\% (AUROC) and 14.08\% (Precision) on NDC\_class. These improvements demonstrate the effectiveness of the SEHP design.

\begin{table}[thbp]
    \centering
    \caption{Results of hyperedge prediction on large-scale datasets on baselines and proposed methods}
    \resizebox{0.35 \textwidth}{!}{%
    \begin{tabular}{c|cc|cc}
        \hline
        Dataset & \multicolumn{2}{c|}{Recipe100k} & \multicolumn{2}{c}{Recipe200k} \\ \hline
        Matrices & AUROC & Precision & AUROC & Precision \\ \hline
        HyperGCN & 0.7568 & 0.7260 & 0.5448 & 0.5310 \\ \hdashline[0.5pt/2pt]
        UniGCNII & 0.6503 & 0.6457 & 0.6823 & 0.6498 \\ \hdashline[0.5pt/2pt]
        EDGNN    & 0.7803 & 0.7113 & 0.6470 & 0.5103 \\ \hdashline[0.5pt/2pt]
        HDS      & 0.7604 & 0.7254 & 0.6276 & 0.5068 \\ \hline
        SEHP     & \textbf{0.8859} & \textbf{0.7342} & \textbf{0.9179} & \textbf{0.6734} \\
        \hline
    \end{tabular}
    }
    \label{tab:large_scale}
\end{table}

\begin{table*}[ht]
    \caption{Ablation study results on small datasets (Cora, DBLP, Citeseer, NDC\_class), comparing SEHP model variations across test sets (SNS, MNS, CNS, MIX) using AUROC and Precision metrics.}
    \centering
    \resizebox{0.98 \textwidth}{!}{%
    \begin{tabular}{l|cccccccccc|cccccccccc}
        \toprule
        Dataset  & \multicolumn{10}{c|}{Cora}                                                                                               & \multicolumn{10}{c}{DBLP}                                                                                                \\ \hline
        Matrices & \multicolumn{5}{c|}{AUROC}                                            & \multicolumn{5}{c|}{Precision}                   & \multicolumn{5}{c|}{AUROC}                                            & \multicolumn{5}{c}{Precision}                    \\ \hline
        Test Set & SNS & MNS & CNS & \multicolumn{1}{c|}{Mix} & \multicolumn{1}{c|}{AVE} & SNS & MNS & CNS & \multicolumn{1}{c|}{Mix} & AVE & SNS & MNS & CNS & \multicolumn{1}{c|}{Mix} & \multicolumn{1}{c|}{AVE} & SNS & MNS & CNS & \multicolumn{1}{c|}{Mix} & AVE \\ \hline
        SEHP-gns & 0.944 & 0.919 & 0.808 & \multicolumn{1}{c|}{0.820}    & \multicolumn{1}{c|}{0.873}    & 0.664 & 0.645 & \textbf{0.669} & \multicolumn{1}{c|}{0.643} & \multicolumn{1}{c|}{0.655}  & 0.947 & 0.779 & 0.733 & \multicolumn{1}{c|}{0.819} & \multicolumn{1}{c|}{0.819} &     0.710 & 0.604 & 0.581 & \multicolumn{1}{c|}{0.607} & \multicolumn{1}{c}{0.6262}   \\ \hline
        SEHP-epre & 0.954 & 0.928 & 0.828 & \multicolumn{1}{c|}{0.817}    & \multicolumn{1}{c|}{0.882}    & 0.712 & 0.721 & 0.622 & \multicolumn{1}{c|}{0.632} & \multicolumn{1}{c|}{0.672}  & 0.943 & 0.816 & \textbf{0.754} & \multicolumn{1}{c|}{0.840} & \multicolumn{1}{c|}{0.838} &     0.832 & 0.685 & 0.617 & \multicolumn{1}{c|}{0.676} & \multicolumn{1}{c}{0.701}   \\ \hline
        SEHP-None & 0.964 & 0.860 & 0.691 & \multicolumn{1}{c|}{0.771}    & \multicolumn{1}{c|}{0.822}    & \textbf{0.756} & 0.706 & 0.559 & \multicolumn{1}{c|}{0.621} & \multicolumn{1}{c|}{0.660}   & 0.938 & 0.791 & 0.739 & \multicolumn{1}{c|}{0.822} & \multicolumn{1}{c|}{0.823} &     0.743 & 0.500 & 0.597 & \multicolumn{1}{c|}{0.630} & \multicolumn{1}{c}{0.617}  \\ \hdashline[0.5pt/2pt]
        SEHP-Stru & 0.954 & 0.932 & 0.839 & \multicolumn{1}{c|}{0.854}    & \multicolumn{1}{c|}{0.895}    & 0.742 & 0.687 & 0.607 & \multicolumn{1}{c|}{0.653} & \multicolumn{1}{c|}{0.672}  & 0.942 & 0.801 & 0.740 & \multicolumn{1}{c|}{0.831} & \multicolumn{1}{c|}{0.828} &     0.841 & 0.673 & 0.611 & \multicolumn{1}{c|}{0.678} & \multicolumn{1}{c}{0.701}  \\ \hdashline[0.5pt/2pt]
        SEHP-Node & 0.964 & 0.931 & 0.816 & \multicolumn{1}{c|}{0.799}    & \multicolumn{1}{c|}{0.878}    & 0.747 & 0.722 & 0.637 & \multicolumn{1}{c|}{0.622} & \multicolumn{1}{c|}{0.682}  & 0.960 & 0.824 & 0.745 & \multicolumn{1}{c|}{0.846} & \multicolumn{1}{c|}{0.844} &     0.854 & 0.681 & 0.614 & \multicolumn{1}{c|}{0.681} & \multicolumn{1}{c}{0.707}  \\ \hdashline[0.5pt/2pt]
        SEHP & \textbf{0.974} & \textbf{0.951} & \textbf{0.850} & \multicolumn{1}{c|}{\textbf{0.856}}    & \multicolumn{1}{c|}{\textbf{0.908}}    & 0.746 & \textbf{0.774} & 0.638 & \multicolumn{1}{c|}{\textbf{0.654}} & \multicolumn{1}{c|}{\textbf{0.703}}  & \textbf{0.961} & \textbf{0.827} & 0.749 &  \multicolumn{1}{c|}{\textbf{0.847}} &  \multicolumn{1}{c|}{\textbf{0.846}} &     \textbf{0.855} & \textbf{0.690} & \textbf{0.620} & \multicolumn{1}{c|}{\textbf{0.682}} & \multicolumn{1}{c}{\textbf{0.712}}    \\ \hline
    \end{tabular} 
    }
    \resizebox{0.98 \textwidth}{!}{%
    \begin{tabular}{l|cccccccccc|cccccccccc}
        \toprule \toprule
        Dataset  & \multicolumn{10}{c|}{Citeseer}                                                                                             & \multicolumn{10}{c}{NDC\_class}                                                                                                \\ \hline
        Matrices & \multicolumn{5}{c|}{AUROC}                                            & \multicolumn{5}{c|}{Precision}                   & \multicolumn{5}{c|}{AUROC}                                            & \multicolumn{5}{c}{Precision}                    \\ \hline
        Test Set & SNS & MNS & CNS & \multicolumn{1}{c|}{Mix} & \multicolumn{1}{c|}{AVE} & SNS & MNS & CNS & \multicolumn{1}{c|}{Mix} & AVE & SNS & MNS & CNS & \multicolumn{1}{c|}{Mix} & \multicolumn{1}{c|}{AVE} & SNS & MNS & CNS & \multicolumn{1}{c|}{Mix} & AVE \\ \hline
        SEHP-gns & 0.924 & 0.819 & 0.690 & \multicolumn{1}{c|}{0.820}    & \multicolumn{1}{c|}{0.813}    & 0.721 & 0.703 & 0.610 & \multicolumn{1}{c|}{0.666} & \multicolumn{1}{c|}{0.675}  & 0.957 & 0.803 & 0.731 & \multicolumn{1}{c|}{0.837}    & \multicolumn{1}{c|}{0.832}    & 0.778 & 0.662 & 0.683 & \multicolumn{1}{c|}{0.691} & \multicolumn{1}{c}{0.704}  \\ \hline
        SEHP-epre & 0.941 & 0.822 & 0.694 & \multicolumn{1}{c|}{0.837}    & \multicolumn{1}{c|}{0.824}    & \textbf{0.772} & 0.683 & 0.591 & \multicolumn{1}{c|}{0.675} & \multicolumn{1}{c|}{0.680}  & 0.948 & 0.854 & 0.761 & \multicolumn{1}{c|}{0.842}    & \multicolumn{1}{c|}{0.846}    & 0.829 & 0.748 & \textbf{0.685} & \multicolumn{1}{c|}{0.692} & \multicolumn{1}{c}{0.738}  \\ \hline
        SEHP-None & 0.942 & 0.827 & 0.682 & \multicolumn{1}{c|}{0.828}    & \multicolumn{1}{c|}{0.820}    & 0.735 & 0.705 & 0.603 & \multicolumn{1}{c|}{0.663} & \multicolumn{1}{c|}{0.676}  & 0.949 & 0.855 & 0.718 & \multicolumn{1}{c|}{0.816}    & \multicolumn{1}{c|}{0.834}    & 0.795 & 0.731 & 0.651 & \multicolumn{1}{c|}{0.673} & \multicolumn{1}{c}{0.712}  \\ \hdashline[0.5pt/2pt]
        SEHP-Stru & 0.931 & 0.834 & \textbf{0.702} & \multicolumn{1}{c|}{0.839}    & \multicolumn{1}{c|}{0.827}    & 0.753 & 0.697 & 0.622 & \multicolumn{1}{c|}{0.665} & \multicolumn{1}{c|}{0.684}  & 0.950 & \textbf{0.871} & 0.721 & \multicolumn{1}{c|}{0.822}    & \multicolumn{1}{c|}{0.841}    & 0.801 & 0.732 & 0.646 & \multicolumn{1}{c|}{0.673} & \multicolumn{1}{c}{0.713}  \\ \hdashline[0.5pt/2pt]
        SEHP-Node & 0.941 & 0.836 & 0.694 & \multicolumn{1}{c|}{0.832}    & \multicolumn{1}{c|}{0.826}    & 0.729 & 0.712 & 0.616 & \multicolumn{1}{c|}{0.650} & \multicolumn{1}{c|}{0.677}  & 0.950 & 0.846 & 0.727 & \multicolumn{1}{c|}{0.827}    & \multicolumn{1}{c|}{0.838}    & 0.796 & 0.733 & 0.659 & \multicolumn{1}{c|}{0.681} & \multicolumn{1}{c}{0.717}  \\ \hdashline[0.5pt/2pt]
        SEHP & \textbf{0.948} & \textbf{0.841} & 0.700 & \multicolumn{1}{c|}{\textbf{0.841}}    & \multicolumn{1}{c|}{\textbf{0.832}}    & 0.763 & \textbf{0.719} & \textbf{0.629} & \multicolumn{1}{c|}{\textbf{0.676}} & \multicolumn{1}{c|}{\textbf{0.686}}  & \textbf{0.982} & 0.861 & \textbf{0.788} & \multicolumn{1}{c|}{\textbf{0.873}}    & \multicolumn{1}{c|}{\textbf{0.876}}    & \textbf{0.850} & \textbf{0.771} & 0.665 & \multicolumn{1}{c|}{\textbf{0.721}} & \multicolumn{1}{c}{\textbf{0.752}}  \\ \bottomrule
    \end{tabular}
    \label{tab:hyperedge_prediction_ablation}
    }
\end{table*}

We also observed that all models performed best on SNS, followed by MNS, with the lowest performance on CNS. This is because MNS and CNS incorporate structural information when generating negative samples for test data, making them more challenging than SNS. Notably, while SEHP follows this trend, the drop in its performance from SNS to MNS and CNS is less severe compared to other models. For example, on DBLP, EDGNN's AUROC dropped from 0.809 on SNS to 0.367 and 0.391 on MNS and CNS, respectively, whereas SEHP maintained higher scores of 0.961, 0.827, and 0.749 across the same conditions. This indicates that SEHP is more robust in handling complex tasks due to its diffusion model, which generates challenging negative samples, helping the model differentiate them more effectively during training.

Furthermore, hyperedge prediction-focused methods like HyperSAGNN, NHP, AHP, and SEHP outperformed hypergraph embedding methods (HyperGCN, UniGCNII, EDGNN, and HDS). On Cora, DBLP, Citeseer, and NDC\_class datasets, hypergraph embedding methods had mean AUROC scores of 0.475, 0.497, 0.477, and 0.576, respectively, while hyperedge-focused models achieved higher scores of 0.632, 0.753, 0.621, and 0.761. 
Precision results showed a similar trend, with hyperedge prediction-focused methods  outperforming hypergraph embedding methods across all datasets. This performance gap is because hypergraph embedding methods are not specifically designed for hyperedge prediction, making them less effective in capturing hyperedge structures. These results emphasize the advantage of using methods tailored for hyperedge prediction.

\subsubsection{Main Results}
\textbf{Compared with SOTA on Large Datasets.}
\label{sec:Compared with SOTA on Large Datasets}
We compared our model with hypergraph embedding methods (HyperGCN, UniGCNII, EDGNN, and HDS) as hyperedge prediction-focused methods like AHP and NHP cannot scale to large datasets. AHP’s output layer size depends on the total number of nodes, while NHP requires pre-building two matrices for negative and positive hyperedges, limiting their scalability.

For large datasets, we evaluated using only the SNS test set, as generating MNS and CNS is time-intensive and their effects are already analyzed on smaller datasets. Table~\ref{tab:large_scale} shows the results on Recipe100k and Recipe200k, where SEHP outperformed all baselines in both AUROC and Precision. On Recipe100k, SEHP achieved an AUROC of 0.8859 and Precision of 0.7342, compared to EDGNN's 0.7803 and 0.7113, showing improvements of 13.54\% and 3.22\%. On Recipe200k, SEHP achieved an AUROC of 0.9179 and Precision of 0.6734, compared to UniGCNII's 0.6823 and 0.6498, with improvements of 34.53\% and 3.63\%. These results confirm SEHP's scalability and effectiveness on large datasets. While hypergraph embedding methods remain scalable, they continue to underperform compared to hyperedge prediction-focused methods, as observed in the smaller datasets.


\subsubsection{Ablation Study}
\label{sec:Ablation Study}
To understand the contributions of SEHP's components, we conducted an ablation study. It examines the impact of structural information, the diffusion model's effectiveness, and the effect of the negative hyperedge generation acceleration scheme. The efficiency of the negative hyperedge generation acceleration scheme is discussed in the next section ``Analysis of Training Time''(\ref{sec:traingTimeAnalysis}); here, we focus only on its effectiveness.
\begin{itemize}
    \item \textbf{SEHP}: The complete model, integrating both sub-hypergraph structure injection and node embedding querying.
    \item \textbf{SEHP-gns}: Replaces SEHP's negative generator with a MLP, similar to AHP, without denoising process.
    \item \textbf{SEHP-epre}: Uses generated negative hyperedge representations directly for classification, bypassing node ID extraction (see subsection~\ref{sec:accelerating_training_process} and Figure~\ref{fig:framework_edge_representation}).
    \item \textbf{SEHP-None}: Omits sub-hypergraph structure injection and node embedding querying in the negative generator.
    \item \textbf{SEHP-Stru}: Incorporates only sub-hypergraph structure injection, excluding node embedding querying.
    \item \textbf{SEHP-Node}: Utilizes node embedding querying exclusively, without sub-hypergraph structure injection.
\end{itemize}
\begin{table}[th]
    \centering
    \caption{Ablation study results on large-scale datasets (Recipe100k and Recipe200k). The table presents the AUROC and Precision scores for different SEHP model variations.}
    \label{tab:hyperedge_prediction_ablation_large}
    \resizebox{0.37 \textwidth}{!}{
    \begin{tabular}{lcccc}
        \toprule
        Matrices & \multicolumn{2}{c}{Recipe100k} & \multicolumn{2}{c}{Recipe200k} \\
        \cmidrule(lr){2-3} \cmidrule(lr){4-5}
         & AUROC & Precision & AUROC & Precision \\
        \midrule
        SEHP-gns  & 0.8715 & 0.7083   & 0.9386 & 0.7544    \\ \hline
        SEHP-epre & \textbf{0.8882} & 0.7311   & 0.9522 & 0.7675    \\ \hline
        SEHP-None  & 0.8792 & 0.7152   & 0.9318 & 0.7444    \\ \hdashline[0.5pt/2pt]
        SEHP-Stru  & 0.8838 & 0.7310   & 0.9413 & 0.7740    \\ \hdashline[0.5pt/2pt]
        SEHP-Node  & 0.8811 & 0.7175   & 0.9478 & 0.7502    \\ \hdashline[0.5pt/2pt]
        SEHP  & 0.8859 & \textbf{0.7342}   & \textbf{0.9749} & \textbf{0.7786}    \\        
        \bottomrule
    \end{tabular}
    }
\end{table}
 

\begin{table*}[ht]
    \centering
    \caption{Comparison of execution times per epoch (in seconds) and speed ratios across different datasets.
    }
    \label{tab:training_time}
    \resizebox{0.70 \textwidth}{!}{
    \begin{tabular}{l|rrrr|c|cc}
    \toprule
    Dataset & AHP & HyperSAGNN  & NHP & SEHP & SEHP-epre & Speed Ratios & Speed Ratios  \\
     &  &  &  &  & & (NHP/SEHP-epre) &  (SEHP/SEHP-epre) \\
    \midrule
    Cora         & 0.946 & 0.422 & 0.326 & 2.026 & 0.062 & 5.26 & \textbf{32.68} \\
    Citeseer     & 0.731 & 0.695 & 0.434 & 2.112 & 0.050 & 8.68 & \textbf{42.24} \\
    DBLP         & 21.586 & 8.146 & 4.037 & 50.969 & 0.717 & 5.63 & \textbf{71.09} \\
    NDC\_class   & 0.892 & 0.533 & 0.401 & 2.182 & 0.076 & 5.28 & \textbf{28.71} \\
    Recipe 100k  & - & - & - & 26.253 & 1.189 & - & \textbf{22.08} \\
    Recipe 200k  & - & - & - & 40.448 & 1.953 & - & \textbf{20.71} \\
    \bottomrule
    \end{tabular}
    }
\end{table*}

The results in Tables~\ref{tab:hyperedge_prediction_ablation} and~\ref{tab:hyperedge_prediction_ablation_large} show that SEHP consistently outperforms its ablated versions, achieving the highest AUROC and Precision across all datasets. For AUROC, SEHP scores 0.908 (Cora), 0.846 (DBLP), 0.832 (Citeseer), and 0.876 (NDC\_class). SEHP-Node and SEHP-Stru closely follow, highlighting the contributions of node embedding querying and sub-hypergraph structure. In terms of Precision, SEHP records 0.703 (Cora), 0.712 (DBLP), 0.686 (Citeseer), and 0.752 (NDC\_class), with SEHP-Node and SEHP-Stru again performing well.
SEHP-None, which lacks both structural components, consistently shows the lowest performance, underscoring the importance of integrating these features. SEHP-gns, using a simple MLP similar to AHP, also performs worse than the full SEHP model, demonstrating the effectiveness of the diffusion-based approach. Notably, SEHP-epre shows competitive results by directly generating negative hyperedges from the latent space, bypassing node ID extraction. This validates the reasonableness of the new architecture which avoids the discrete ID generation for the diffusion models by directly performing the negative sampling in the latent space instead of in the discrete ID space. 

Overall, the ablation study confirms the effectiveness of the SEHP design. The full model outperforms ablated versions across most datasets and evaluation metrics, proving that each component effectively enhances overall performance.

\subsubsection{Analysis of Training Time}
\label{sec:traingTimeAnalysis}
We evaluate the per-epoch training time of SEHP and SEHP-epre compared to other hyperedge prediction-focused methods (AHP, HyperSAGNN, NHP) to assess training efficiency and the impact of our acceleration scheme. 
SEHP-epre significantly improves efficiency (Table~\ref{tab:training_time}). On Recipe100k, it trains in 1.19 seconds per epoch, 22 times faster than SEHP's 26.25 seconds. On Recipe200k, SEHP-epre (1.95 seconds) is 21 times faster than SEHP (40.45 seconds), effectively accelerating training.
AHP, NHP, and HyperSAGNN cannot run on large hypergraphs (Recipe100k and Recipe200k) due to scalability issues, so their times are presented only on the four smaller datasets (Table~\ref{tab:training_time}). SEHP-epre is approximately 6.21 times faster than the second-fastest model (NHP) across these datasets. SEHP-epre consistently outperforms all baselines in effectiveness (Tables~\ref{tab:hyperedge_prediction_ablation}, Table~\ref{tab:hyperedge_prediction_baselines}), demonstrating faster and better performance, with scalability to large datasets.


\section{Conclusion}
\label{sec:conclusion}



We introduce SEHP, a scalable framework for negative sample generation in hyperedge prediction. SEHP leverages a conditional diffusion model to iteratively generate and refine negative hyperedges, enhancing performance. To our knowledge, SEHP is the first to apply diffusion models in hyperedge prediction. By incorporating hypergraph structure and sampling sub-hypergraphs for batch training, SEHP addresses both quality and scalability challenges. Our architecture directly utilizes latent space representations, eliminating the need for discrete ID generation and achieving a 20 to 71-fold speedup across six datasets with minimal performance loss. Experiments on real-world datasets demonstrate that SEHP outperforms state-of-the-art methods in both accuracy and scalability. Ablation studies confirm that the diffusion-based generator and hypergraph structure integration are essential for performance. SEHP offers a scalable solution for large hypergraphs and opens new possibilities for applying diffusion models in hypergraph tasks.


\newpage
\balance
\bibliographystyle{ACM-Reference-Format}
\bibliography{sample-base,my}
\appendix

\onecolumn
\section{Additional Figures and Tables}
\begin{figure*}[bh]
    \centerline{\includegraphics[width=0.75\textwidth]{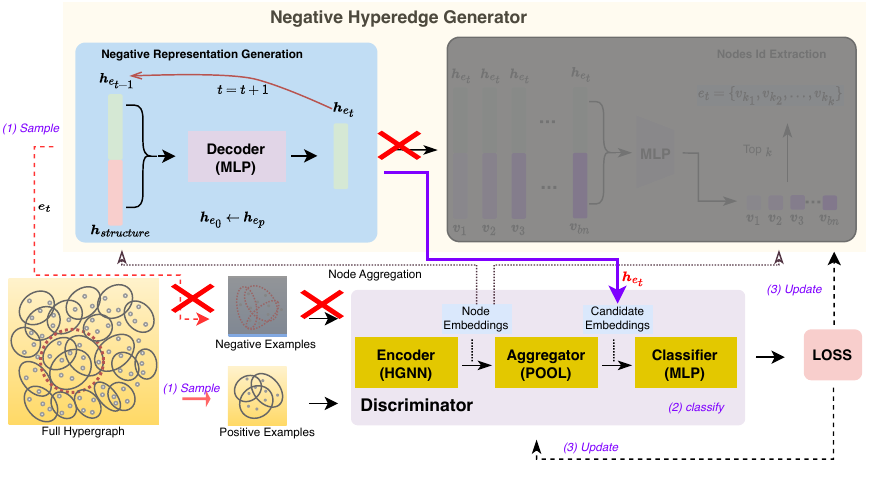}}
    \caption{Framework of the conditional diffusion-based negative hyperedge generation for hyperedge prediction using generated hyperedge embeddings directly. The figure illustrates the modified approach where node ID extraction is bypassed, allowing the generated negative hyperedge representation to be used directly for classification. This streamlined process enhances efficiency by leveraging the continuous embedding from the diffusion model without discrete node selection.}
    \Description{Framework of the conditional diffusion-based negative hyperedge generation for hyperedge prediction using generated hyperedge embeddings directly. The figure illustrates the modified approach where node ID extraction is bypassed, allowing the generated negative hyperedge representation to be used directly for classification. This streamlined process enhances efficiency by leveraging the continuous embedding from the diffusion model without discrete node selection.}
    \label{fig:framework_edge_representation}
\end{figure*}

\begin{table*}[ht]
    \caption{Summary statistics for all datasets. `Node' indicates the number of nodes, `HEdge' represents the number of hyperedges, `HEdge Connect' shows the average number of nodes per hyperedge, and `Node Connect' indicates the average number of hyperedges to which a node belongs.}
    \centering
    \resizebox{0.43 \textwidth}{!}{%
        \begin{tabular}{ccccc}
            \toprule
            Dataset        & Node & HEdge & HEdge Connect & Node Connect \\ \hline
            Cora          & 2,388    & 970     & 4.48    & 1.82     \\
            DBLP          & 41,302   & 20,865  & 4.61    & 2.33     \\
            Citeseer      & 1,458    & 1,004   & 3.25    & 2.24     \\
            NDC\_class  & 1,161    & 1,047   & 6.11    & 5.51     \\
            Recipe100k    & 100,896  & 11,822  & 93.81   & 10.99    \\
            Recipe200k    & 240,094  & 18,049  & 244.54  & 18.38    \\ \bottomrule
        \end{tabular}%
    }
    \label{tab:dataset_summary}
    \end{table*}  


\end{document}